\documentclass[reprint,aps,prl,superscriptaddress,twocolumn,floatfix]{revtex4}
\usepackage{graphicx}
\usepackage{amsmath}
\usepackage{bm}
\usepackage{color}

\begin{document}

\title{Slow dynamics in a quasi-two-dimensional binary complex plasma}

\author{Cheng-Ran Du}
\email{chengran.du@dhu.edu.cn}
\affiliation{College of Science, Donghua University, Shanghai 201620, PR China}
\author{Vladimir Nosenko}
\author{Hubertus M. Thomas}
\affiliation{Institut f\"ur Materialphysik im Weltraum, Deutsches Zentrum f\"ur Luft- und Raumfahrt, 82234 We{\ss}ling, Germany}
\author{Yi-Fei Lin}
\affiliation{College of Science, Donghua University, Shanghai 201620, PR China}
\author{Gregor E. Morfill}
\affiliation{BMSTU Centre for Plasma Science and Technology, Moscow, Russia}
\author{Alexei V. Ivlev}
\email{ivlev@mpe.mpg.de} \affiliation{Max Plank Institute for Extraterrestrial Physics, Garching 85748, Germany}

\begin{abstract}
Slow dynamics in an amorphous quasi-two-dimensional complex plasma, comprised of microparticles of two different sizes, was studied experimentally. The motion of individual particles was observed using video microscopy, and the self-part of the intermediate scattering function as well as the mean-squared particle displacement was calculated. The long-time structural relaxation reveals the characteristic behavior near the glass transition. Our results suggest that binary complex plasmas can be an excellent model system to study slow dynamics in classical supercooled fluids.
\end{abstract}

\maketitle

When a fluid is quenched by cooling or compression, it may either crystallize or remain in an amorphous state, depending on the complexity of the fluid and the quenching depth. Such fluids are said to be ``supercooled'' when they are still able to equilibrate in the experimental time window, exhibiting a slow structural relaxation caused by rare rearrangement of atoms or molecules \cite{Debenedetti:2001,Berthier:2011}. Otherwise, they become dynamically arrested and undergo the glass transition. Understanding the mechanisms governing the slow dynamics and the approach to the glass transition is a fundamental problem of classical condensed matter physics \cite{Franosch:1997,Voigtmann:2004,Cavagna:2009,Flenner:2015}.

The relaxation timescale in molecular glasses is $\sim14$ orders of magnitude longer than that in high-temperature liquids \cite{Debenedetti:2001}, which makes the glass transition inaccessible for up-to-date numerical simulations. For this reason, model soft-matter systems play a crucial role in the study of slow dynamics \cite{Cipelletti:2005,Ivlev:Book}. Among these, colloidal suspensions \cite{vanMegen:1993,Weeks:2000,Mattsson:2009} and granular matter \cite{Richard:2005,Xia:2015} have drawn particular attention. As equilibrium strongly damped systems, colloidal suspensions exhibit Brownian dynamics \cite{Foffi:2003}, while essentially non-equilibrium granular matter obeys Newtonian microscopic dynamics \cite{Sperl:2010}, with the dissipation introduced in mutual particle collisions \cite{Kranz:2010,Sperl:2012,Berthier:2013}. Due to their reasonable experimental timescales and straightforward diagnostic methods, both systems provide excellent conditions for particle-resolved studies of slow dynamics.

Complex plasmas, composed of a weakly ionized gas and charged microparticles, represent the plasma state of soft matter \cite{Ivlev:Book}. They have several remarkable features distinguishing them from other soft-matter systems \cite{Fortov:2005,Morfill:2009}. First, since the background gas is dilute, the short-time particle dynamics in strongly coupled complex plasmas is virtually undamped, which provides a direct analogy to regular liquids and solids in terms of the atomistic dynamics. Second, most notable, the interparticle interactions generally violate the action-reaction symmetry \cite{Schweigert:1996}. In stable {\it binary} (quasi-2D) complex plasmas \cite{Ivlev:2017} the non-reciprocal interactions lead to a {\it dynamical equilibrium} \cite{Ivlev:2015}, where different particle species have distinct kinetic temperatures. For a special class of interactions with a constant non-reciprocity, the dynamical equilibrium is {\it detailed}. This latter remarkable property of quasi-2D complex plasmas allows us to employ standard methods of equilibrium statistical mechanics for their description.

In this Letter, we report on the first dedicated study of slow dynamics in quasi-2D complex plasmas. A binary mixture of microparticles was used to suppress crystallization and form an amorphous state. To describe the collective dynamics and the structural relaxation, we measured the mean-squared particle displacement (MSD) and the self part of the intermediate scattering function (ISF). The evolution of MSD exhibits a crossover from the short-time ballistic dynamics to a transient sub-diffusive behavior determined by collective interactions. The long-time decay of ISF reveals characteristic features of supercooled fluids approaching the glassy state. The presented results demonstrate complementary advantages of quasi-2D complex plasmas and point out their remarkable dynamical properties with respect to other soft-matter systems.

\begin{figure}
	\includegraphics[width=8 cm]{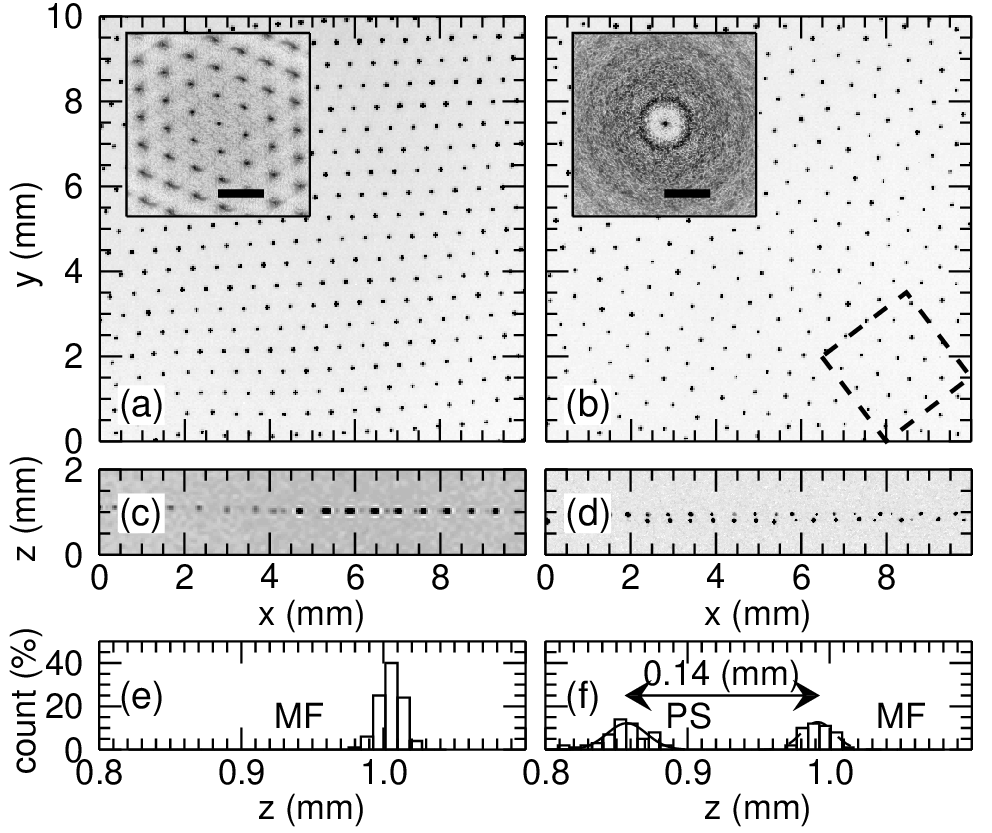}
	\caption{Comparison of a 2D plasma crystal (left panels) with a quasi-2D amorphous state (right panels), obtained for equivalent plasma conditions. The top views and the corresponding Fourier-transformed images in the insets (representative of the static structure factor) demonstrate that monodisperse MF particles form a monocrystal with a triangular lattice (a), while a binary mixture of MF and PS particles (with the mixing ratio about $1:1$) shows neither translational nor orientational long-range order (b). The scale bars represent $20$~mm$^{-1}$. A square lattice domain is highlighted by the dashed rectangle in (b). The side views and the corresponding height histograms for the crystal (c,e) and the binary mixture (d,f) reveal differences in the levitation heights ($z$-coordinates) of individual particles in the two cases.}
	\label{fig1}
\end{figure}

{\it Experiment.} The experiment was performed in a modified Gaseous Electronics Conference (GEC) rf reference cell \cite{Feng:2008,Couedel:2010,Nosenko:2009,Du:2012}. The plasma was produced with a capacitively coupled rf discharge in argon at a pressure of $0.66$~Pa. The negatively-charged particles were levitated in the plasma sheath above the bottom electrode, where the gravity force is balanced by the electric force. Individual particles were illuminated by a laser sheet from the side and their motions were recorded by a CMOS camera from the top. To suppress crystallization, we used a mixture \cite{Smith:2008,Hartmann:2009,Kalman:2013,Wieben:2017} of Melamine Formaldehyde (MF) and Polystyrene (PS) microparticles with diameters of $9.19~\mu$m and $11.36~\mu$m, respectively, suspended at almost the same height. The discharge power was a control parameter, to quench the binary complex plasma. Unlike a 2D suspension of monodisperse particles [Fig.~\ref{fig1}~(a)], the quasi-2D binary system was amorphous [Fig.~\ref{fig1}~(b)]. The particle suspension slowly rotated \cite{Konopka:2000,Carstensen:2009}, which may have been induced by the inhomogeneity of the laser illumination. To mitigate this problem, we placed two aluminium bars on the rf electrode, parallel to each other and separated by 9~cm. As a result, the angular velocity was drastically reduced to $\Omega\sim10^{-3}$~rad/s; see S.1 of the Supplemental Material \cite{supplemental} for more details on the experimental procedure.

{\it Results and analysis.} The crystalline and amorphous complex plasmas shown in Fig.~\ref{fig1} had practically the same areal densities, with the mean horizontal interparticle distance of $\Delta \simeq 0.55$~mm measured from the first peak of the respective pair correlation function, (see Fig.~2 in the Supplemental Material \cite{supplemental}).  The density inhomogeneity was within 1\%. The qualitative difference between the crystalline and amorphous states is conveniently illustrated with the static structural analysis. By applying a 2D Fast Fourier Transformation (FFT) on top-view snapshots, panels (a) and (b), we obtained the respective diffraction patterns, plotted in the insets in log scale and representing the static structure factor of the studied systems. One can see that the diffraction pattern of the binary mixture exhibits isotropic concentric rings, typical for amorphous materials.

The side view in Fig.~\ref{fig1} shows that MF particles in the crystal were levitated at (practically) the same height [panels (c) and (e)], while MF particles in a binary mixture were suspended slightly higher than the PS particles [panels (d) and (f)]. The height difference of $0.14$~mm, determined from a Gaussian fit of the height histograms, is about a quarter of $\Delta$. The interparticle interactions in this case become essentially non-reciprocal due to the presence of plasma wakes \cite{Schweigert:1996,Lampe:2000,Fortov:2005,Kompaneets:2007,Morfill:2009}, and a binary mixture tends to a dynamical equilibrium, where the upper particles have a higher temperature of the horizontal motion than the lower particles \cite{Ivlev:2015}. In our experiment, the kinetic temperature was determined separately for the upper and lower particles, from a Maxwellian fit of the corresponding velocity distributions. In agreement with the theoretical predictions \cite{Ivlev:2015}, the resulting temperature of the upper particles, $T_{\rm MF}\simeq 1100$~K, was substantially higher than the temperature of the lower particles, $T_{\rm PS}\simeq930$~K. The particle charges, $Q_{\rm MF} \simeq 13000e$ and $Q_{\rm PS} \simeq 16000e$ (with an uncertainty of $30\%$), were deduced from the phonon spectra of the corresponding {\it crystalline suspensions} under equivalent discharge conditions. Simultaneously, these measurements yielded the effective plasma screening length of $\lambda\simeq 0.4$~mm for the presented example.

The thermodynamic state of a charged system is characterized by the coupling and screening parameters \cite{Ivlev:Book,Hartmann:2005}. For a binary mixture, the relevant coupling parameter is defined as $\Gamma= Q_{\rm MF}Q_{\rm PS}\sqrt{n}/(k_{\rm B}\overline{T}$) with $\overline{T}=(T_{\rm MF}+T_{\rm PS})/2$, and the screening parameter is $\kappa= 1/(\lambda\sqrt{n})$. Based on the measured values, we obtained $\Gamma \simeq 6000$ (with an uncertainty of $45\%$), and $\kappa\simeq 1.5$ (with an uncertainty of $30\%$).

\begin{figure}[!ht]
    \includegraphics[width=8 cm]{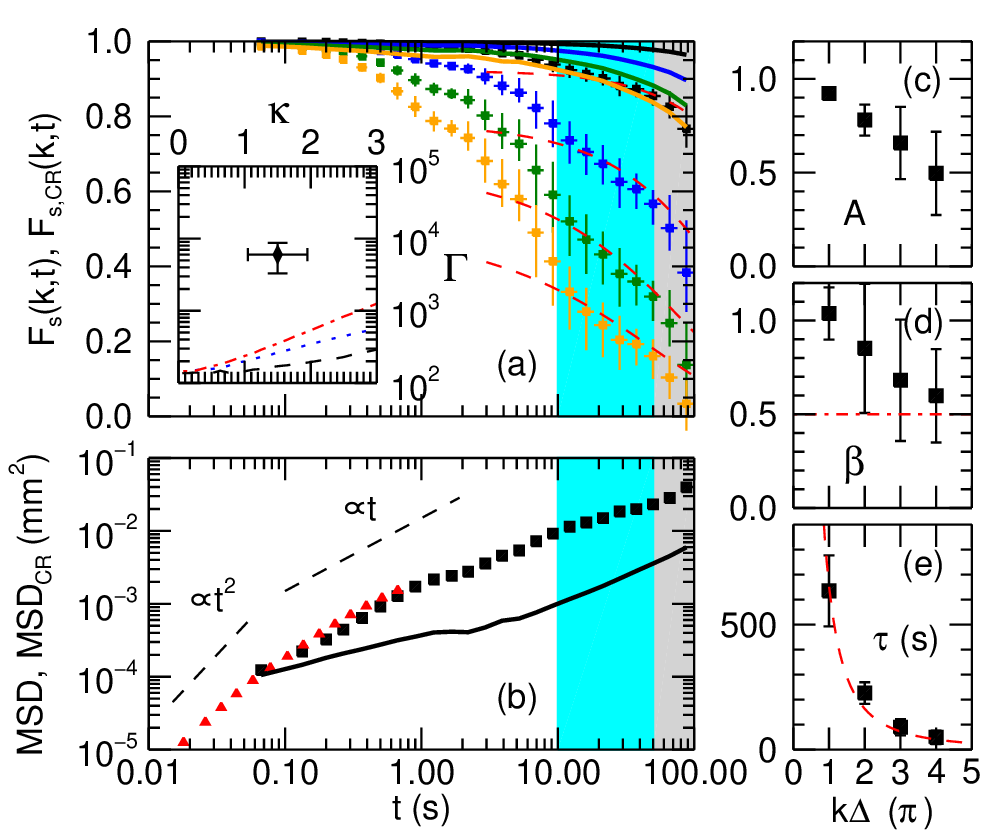}
    \caption{Structural relaxation in a quasi-2D amorphous complex plasma. The results are for the experiment shown in the
    right panels of Fig.~\ref{fig1}. (a) The self-part of the ISF [$F_{\rm s}(k,t)$, squares]  and the cage-relative ISF [$F_{\rm s,CR}(k,t)$, solid lines],  plotted for $k\Delta=\pi$ (black), $2\pi$ (blue), $3\pi$ (green), and $4\pi$ (yellow). (b) MSD measured with long-distance microscope (red triangles) and using a video microscopy technique with micro lens (black squares); the black solid line shows MSD$_{\rm CR}$ measured with the latter technique. The vertical grey stripe marks a gradual crossover to a ``forced'' relaxation induced by a slow rotation and the cyan stripe marks the fit range. The inset shows the glass transition lines derived from MCT \cite{Yazdi:2015} for the Yukawa potential (dashed line) and the Kompaneets potential \cite{Kompaneets:2007} (dotted and dash-dotted lines, representing collision parameter $\zeta=0.5$ and $0.25$, respectively). The symbol above the transition lines shows $\Gamma$ and $\kappa$  deduced for our experiment with errors, indicating that thermodynamically the system forms a glass. Fitting the long-time asymptote of ISF (for a given $k$) with the stretched-exponential law [dashed lines in (a)] yields: the amplitude factor $A$ (c), the stretching exponent $\beta$ (d), and the timescale $\tau$ [(e) the dashed line demonstrates a $1/k^2$ fit]; see S.3 of the Supplemental Material \cite{supplemental} for details.}
    \label{fig2}
\end{figure}

The structural relaxation is generally quantified by the density-density correlation function in ${\bf k}$-space, $F({\bf k},t)$, which is the Fourier-transformation of the van Hove correlation function \cite{Hansen:book}, commonly referred to as ISF. For practical purposes, it is convenient to use the self-part of ISF, $F_{\rm s}({\bf k},t) = N^{-1}\langle\sum_i^N\exp [-i{\bf k}\cdot \Delta{\bf r}_i(t)]\rangle$ with $\Delta{\bf r}_i(t)={\bf r}_i(t+t_0) - {\bf r}_i(t_0)$, describing the evolution of single-particle correlations \cite{Fuchs:1998,Voigtmann:2004,Bayer:2007}. Here, ${\bf r}_i(t)$ is the position of the particle $i$ at the moment $t$, and $\langle\ldots\rangle$ denotes averaging over $t_0$. For the analysis, we selected a region of interest (ROI) such that the ``rattlers'' (a few visibly oscillating irregular particles) and their nearest neighborhood were removed. The structural relaxation in unstressed amorphous materials does not depend on the orientation of the wave vector ${\bf k}$, so here we also averaged over the orientation. The calculated $F_{\rm s}(k,t)$ are shown in Fig.~\ref{fig2}.

The stretched-exponential (Kohlrausch) law \cite{Franosch:1997,Fuchs:1992,Fuchs:1994,Hansen:book,Voigtmann:2004,Feng:2010b},
$F_s(k,t)\simeq A(k)\exp\{-[ t/\tau(k) ]^{\beta(k)}\}$, usually provides a good fit for the long-time asymptote of ISF, the so-called alpha-relaxation. The outcome of the fit is shown in Fig.~\ref{fig2}~(a) by the dashed lines. The law is determined by three parameters: the amplitude factor $A(k)$, the timescale of the alpha-relaxation $\tau(k)$, and the stretching exponent $\beta(k)<1$. Selecting a time domain appropriate for the fit is generally not an easy task \cite{Fuchs:1992,Bartsch:1992,Phillips:1996,Voigtmann:2004} -- an overlap with the transient beta-relaxation should be avoided, which imposes the lower time bound for the fit. In our experiment, we have an additional constraint, associated with a slow rotation of the particle suspension: Although we were able to reduce the angular velocity down to $\Omega\sim10^{-3}$~rad/s, there still has been a profound effect caused by this rotational shear \cite{Yamamoto:1998} at sufficiently long times, where the accumulated strain $2\Omega t$ exceeds a certain critical value. According to Zausch~{\it et~al.}~\cite{Zausch:2008}, the onset of plastic deformations in glassy systems (upon a simple stress) is expected when a strain exceeds a critical value of $\sim10^{-1}$. Using this as a guide, we estimate the upper time limit as $t\sim10^2$~s. Figure~\ref{fig2}~(a) shows that the measured ISF indeed starts falling off rapidly in the time range between 50--100~s, indicating a crossover from the generic alpha-relaxation to the rotation-induced decay.

The Kohlrausch amplitude $A(k)$ and the stretching exponents $\beta(k)$ are plotted in Fig.~\ref{fig2}~(c,d). In agreement
with the mode coupling theory (MCT) of the fluid-glass transition \cite{Fuchs:1992,Voigtmann:2004,Gotze:book}, both $A(k)$ and $\beta(k)$ tend to unity for $k\to0$ and decrease monotonically at large $k$. The timescale of the alpha-relaxation, also obtained from the Kohlrausch fit and shown in Fig.~\ref{fig2}~(e), closely follows a $\tau(k)\propto 1/k^2$ dependence. Such scaling is predicted by MCT at small $k$, while for $k\rightarrow \infty$ it should change to $\tau(k)\propto 1/k^{1/b}$, where $b$ is the asymptotic value of $\beta(k)$ (the von Schweidler exponent) \cite{Franosch:1997,Voigtmann:2004,Bayer:2007,Reis:2007}. As the value of $b$ is expected to be close to $\simeq0.5$, the measured behavior of $\tau(k)$ is in good agreement with the theory \cite{Fuchs:1992,Fuchs:1993}.

Recently, Yazdi~{\it et~al.}~\cite{Yazdi:2015} employed MCT to calculate the idealized glass transition lines for 2D complex plasmas, using the Yukawa and Kompaneets \cite{Kompaneets:2007} potentials for the interparticle interactions. These results, derived for a {\it monodisperse} system, are depicted in the inset of Fig.~\ref{fig2}~(a) where the transition lines are plotted in the $(\Gamma,\kappa)$ plane. Note that for the Kompaneets potential, we plot the transition lines for two typical values of collision parameter $\zeta$ \cite{Kompaneets:2007}  (which is the ratio of a field-induced screening length to the ion-neutral mean free path). This potential provides more realistic description of the interactions in 2D complex plasmas. The values of $\Gamma$ and $\kappa$ deduced for our experiment fall substantially above the transition lines regardless of the model, which indicates that thermodynamically the system formed a glass.

It is instructive to complement the analysis by calculating MSD$=N^{-1}\langle\sum_i^N|\Delta{\bf r}_i(t)|^2\rangle$, which is directly related to ISF in the limit $k\rightarrow0$ \cite{Voigtmann:2004}. Figure~\ref{fig2}(b) shows that MSD exhibits a ballistic behavior $\propto t^2$ for $t\lesssim0.1$~s, i.e., the short-time in-cage motion obeys Newtonian dynamics. In fact, interaction of particles with gas becomes important for $\nu t\gtrsim1$, where $\nu$ is the damping rate due to gas friction \cite{Ivlev:Book} ($\nu_{\rm MF} \simeq 0.8$~s$^{-1}$ and $\nu_{\rm PS} \simeq 0.9$~s$^{-1}$ for our conditions). Thus, up to $t\sim1$~s the measured MSD reflects the generic behavior occurring in molecular supercooled liquids. The behavior becomes substantially sub-diffusive by that time, indicating the onset of transient beta-relaxation, with a gradual transition to the alpha-relaxation regime observed in Fig.~\ref{fig2}~(a) at $t\gtrsim10$~s.

It has been recently discovered that Mermin-Wagner fluctuations induce a significant translational motion in a 2D system \cite{Vivek:2017,Illing:2017}. The resulted collective motion can be subtracted by measuring the cage-relative ISF and MSD, as  $F_{\rm s,CR}({\bf k},t) = N^{-1}\langle\sum_i^N\exp [-i{\bf k}\cdot \Delta{\bf r}_{{\rm CR},i}(t)]\rangle$ and MSD$_{\rm CR}=N^{-1}\langle\sum_i^N|\Delta{\bf r}_{{\rm CR},i}(t)|^2\rangle$, respectively, where $\Delta{\bf r}_{{\rm CR},i}(t)=\Delta{\bf r}_i(t) -N_n^{-1}\sum_{j}^{N_n}\Delta{\bf r}_j(t)$. Here $j$ denotes nearest neighbors of the particle $i$ at initial time $t_0$, and the sum is over all neighbors $N_n$. As shown in Fig.~\ref{fig2}(a,b), the cage-relative relaxation is much slower, indicating that the fluctuations are significant. The measured difference between $F_{\rm s,CR}(k,t)$ and $F_{\rm s}(k,t)$ (averaged over orientations of ${\bf k}$) is substantially larger than that observed in colloids, since the charged particles in a plasma interact via much softer interactions, allowing stronger fluctuations. However, we point out that the shear stress may enhance the collective motion too, and therefore play a role in the discrepancy between $F_{\rm s,CR}(k,t)$ and $F_{\rm s}(k,t)$ (see supplemental material \cite{supplemental}). This requires further careful investigations.

\begin{figure}[!ht]
    \includegraphics[width=8 cm]{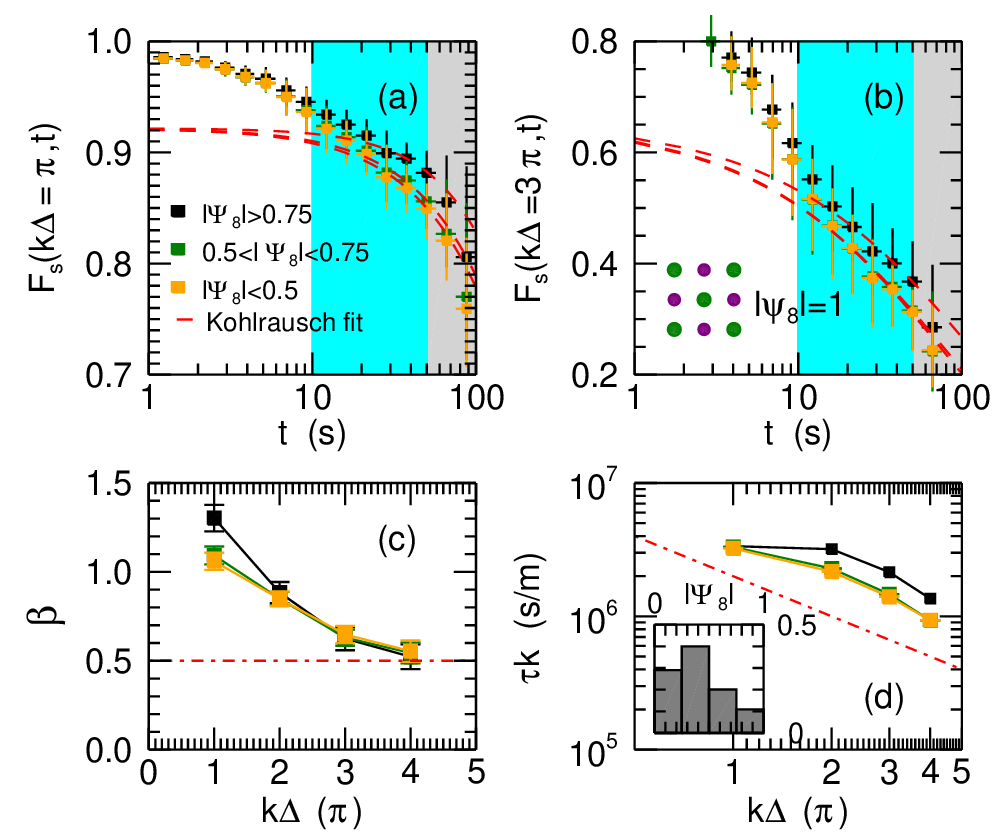}
    \caption{Effect of the local structure on the relaxation dynamics. The self-part of the ISF for particles with different local order parameter $\Psi_8$, plotted for $k\Delta=\pi$ (a) and $3\pi$ (b). k-dependence of  the stretching exponent $\beta$ (c), and the timescale $\tau$ of  alpha-relaxation (d). A perfect square lattice domain ($|\Psi_8|=1$) is depicted in (b). The inset in (d) shows the distribution of $|\Psi_8|$.}
    \label{fig3}
\end{figure}

Fig.~\ref{fig3}(a,b) shows that the dynamical relaxation also depends on the local structure  \cite{wu:2018}. In quasi-2D complex plasmas, square lattice domains of limited size are embedded in amorphous structure \cite{huang:2019,Assoud:2009}. Such structure can be quantified by the local order parameter $\Psi_8=\frac{1}{8}\sum_j e^{i8\theta_j}$, where we only consider eight nearest neighbors and $\theta_j$ is the angle between ${{\bf r}}_j-{{\bf r}}$ and the $x$~axis. The fit of ISF by Kohlrausch law shows that for small $k$, the value of $\beta$ for square domains exceeds unity, implying a compressed exponential relaxation, see Fig.~\ref{fig3}(c). As the majority of the particles in amorphous state show diffusive motion, the particles in the lattice exhibit ballistic motion, represented by the plateau of the product $\tau k$ for small $k$, see Fig.~\ref{fig3}(d). Our results for a quasi-2D system complement the recent numerical simulation in 3D metallic glass-forming melt, where clusters of icosahedra exhibit solid-like compressed exponential relaxation \cite{wu:2018}.

\begin{figure}[!ht]
	\includegraphics[width=8 cm]{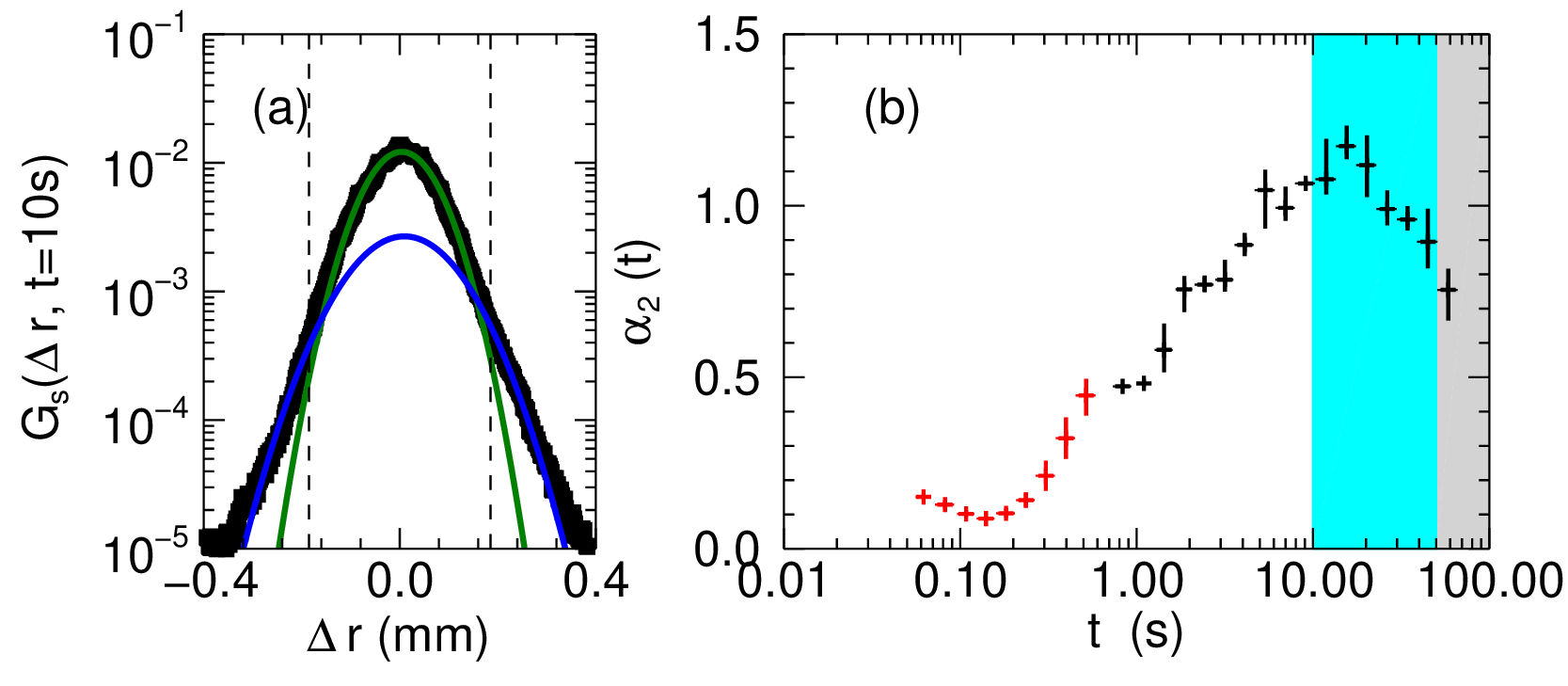}
	\caption{Self-part of the van Hove correlation function, $G_s({\Delta}r,t)$, at a lag time of $10$~s (a) and evolution of the non-Gaussian parameter, $\alpha_2(t)$ (b). The green and blue solid lines in (a) are the Gaussian fits for the slowest $95$\% and fastest $5$\% of particles, respectively. The red and black symbols in (b) represent data obtained with long-distance microscope and with micro lens, respectively [see Fig.~\ref{fig2}(b)]. }
	\label{fig4}
\end{figure}

The dynamical heterogeneity is closely related to the kinetic slowing of alpha-relaxation in supercooled liquids. It can be manifested as a non-Gaussian behavior of the self-part of the van Hove correlation function, $G_s({\Delta}r,t)=N^{-1} \langle\sum_i^N\delta [{\Delta}r-{\Delta}r_i(t)]\rangle$, which is essentially a probability distribution of particle displacement and is Gaussian for purely diffusive particles \cite{Weeks:2000,Kegel:2000,wang:2018}. As shown in Fig.~\ref{fig4}~(a), the distribution is broader at time scales of the alpha-relaxation, as expected for supercooled liquids. The lowest-order deviation of $G_s$ from a Gaussian is quantified by $\alpha_2(t)=\langle{\Delta}r(t)^4\rangle/[(1+d/2)(\langle{\Delta}r(t)^2\rangle^2)]-1$, where $d$ is the dimension of the system. Indeed, Fig.~\ref{fig4}~(b) shows that $\alpha_2(t)$ exhibits a peak at $t\approx10$~s, corresponding to the transition to the alpha-relaxation \cite{Weeks:2000,wang:2018}. Besides, the presence of the square lattice domains may also contribute to the heterogeneity.

{\it Conclusion and outlook.} This Letter reports on the first systematic attempt to experimentally investigate glassy dynamics with complex plasmas. The presented results could be exceptionally important for particle-resolved studies of slow dynamics in future, as complex plasmas are dynamically complementary to other soft-matter systems used for such investigations. We showed that the in-cage motion of individual particles remains virtually undamped (Newtonian), which enables modeling of molecular glasses at timescales up to the transient regime of the beta-relaxation. A crossover to the fully damped Brownian dynamics at longer timescales allows matching with glassy behavior observed in colloidal dispersions \cite{Ivlev:Book}. Furthermore, binary quasi-2D complex plasmas open up a unique opportunity to study fluids with distinct temperatures for different species \cite{Ivlev:2015}, where the temperature mismatch is controlled by the vertical levitation gap. Slow dynamics and glass transitions in such systems may reveal new facets inaccessible to ``regular'' simple liquids.

Our studies have also identified critical issues that need to be resolved to improve the quality of future experiments. The most important is to suppress the global rotation of the suspension: this process terminates the long-time relaxation of ISF as well as a transition to the long-time diffusion. Second, less critical but still important issue are rare ``rattler'' events, sporadically occurring in the field of view and destroying weak correlations in the alpha-relaxation regime. Here, a careful choice of particles used in experiments may be a solution.

\begin{acknowledgments}
The authors acknowledge support from the National Natural Science Foundation of China (NSFC), Grant No. 11405030. We thank J\"urgen Horbach and Eric Weeks for valuable discussions.
\end{acknowledgments}

\end{document}